\documentclass[american,aps,preprint,prd,showpacs,amsfonts,amssymb]{revtex4}
\usepackage[T1]{fontenc}
\usepackage[latin1]{inputenc}
\usepackage{graphics}

\makeatletter

\providecommand{\LyX}{L\kern-.1667em\lower.25em\hbox{Y}\kern-.125emX\@}

 \newenvironment{lyxlist}[1]
   {\begin{list}{}
     {\settowidth{\labelwidth}{#1}
      \setlength{\leftmargin}{\labelwidth}
      \addtolength{\leftmargin}{\labelsep}
      }}
   {\end{list}}

\usepackage[T1]{fontenc}
\usepackage[latin1]{inputenc}
\usepackage{amsmath}
\usepackage{babel}
\usepackage{graphics}


\begin{document}

\newcommand{\bis}{\(\blacktriangleleft\)}

\newcommand{\von}{\(\blacktriangleright\)}

\newcommand{\new}[1]{\von #1\bis}

\preprint{hep-th/0403056}

\preprint{DAMTP-2004-24}


\title{Sectional Curvature Bounds in Gravity:\\Regularisation of the Schwarzschild
Singularity}

\author{Frederic P. Schuller}

\email{fschuller@perimeterinstitute.ca}

\affiliation{Perimeter Institute for Theoretical Physics, Waterloo N2J 2W9, Ontario, Canada}

\author{Mattias N.R. Wohlfarth}

\email{m.n.r.wohlfarth@damtp.cam.ac.uk}

\affiliation{Department of Applied Mathematics and Theoretical Physics, Centre for Mathematical
Sciences, University of Cambridge, Wilberforce Road, Cambridge CB3 0WA, United
Kingdom}

\begin{abstract}
A general geometrical scheme is presented for the construction of novel classical
gravity theories whose solutions obey two-sided bounds on the sectional curvatures
along certain subvarieties of the Grassmannian of two-planes. The motivation
to study sectional curvature bounds comes from their equivalence to bounds on
the acceleration between nearby geodesics. A universal `minimal length' scale
is a necessary ingredient of the construction, and an application of the kinematical
framework to static, spherically symmetric spacetimes shows drastic differences
to the Schwarzschild solution of general relativity by the exclusion of spacelike
singularities. 
\end{abstract}
\maketitle

\section{Introduction}

The history of modifications of Einstein's theory of gravity is a long and winding
one, and its beginnings almost date back to the publication of general relativity
itself \cite{Wey,Edd,Ein,Sch}. A main motivation for all proposed modifications of general relativity
comes from the fact that the spacetime solutions of this theory, even in simple
cases, may contain curvature singularities. The existence of such solutions
creates problems, both mathematical and physical; a famous example is the black
hole information paradoxon which seems to make it impossible to reconcile general
relativity with quantum mechanics \cite{InPa}. The solution of these problems
is usually postponed to an unforeseeable future with the claim that a quantum
theory of gravity would smooth out the singularities. It is indeed sensible
to expect a radically altered short distance behaviour in quantum gravity and
string theory, and this becomes apparent, respectively, in the quantisation
of the area or volume operators in loop quantum gravity, or in the appearance
of higher order curvature corrections to the Einstein-Hilbert action in string
theory, see e.g. \cite{Thi03,Pol98}. But postponing the solution of existing
classical problems to a future quantum theory means accepting to quantise the
already `problematical' classical theory. It might well be the case that this
causes unnecessary additional difficulties. For this reason alone, it is worthwhile
to change the perspective and to reverse the procedure: try to improve on classical
general relativity \textit{before} quantising it.

Modified theories of gravity should conform to a number of criteria, of which
the most important ones are that they should reduce to general relativity in
a limit that covers the range of observations and they should not contain ghost
poles. In the construction of such theories, one can essentially embark on two
different routes. The first is to change the stage, i.e., to enrich or alter
the geometrical structure of the spacetime manifold; examples falling into this
category are theories with non-vanishing torsion, like Einstein-Cartan theory,
e.g. \cite{EC}, or those based on a non-symmetric metric, e.g. \cite{Mof95}.
The second possibility is to provide a new screenplay, i.e., to change the dynamics
of the gravitational field. This happens, for example, in string theory which
provides higher order corrections to the Einstein-Hilbert action while retaining
the Lorentzian structure of the spacetime manifold. The standard way of proceeding,
in either case, is to formulate a consistent new action principle and to hope
that the solutions of the corresponding equations of motion have nicer properties
than the solutions to Einstein's theory. Such constructions, however, can be
at best educated guesses, because underlying geometrical concepts, which are
of utmost importance in any theory of gravity, are very hard, or even impossible,
to identify. In this article, the programme is a different one. Our starting
point is the physical notion of accelerations between nearby geodesics, and
we will investigate the possibility to place bounds on these accelerations.
The kinematical structure of these bounds will be analysed and the results will
be used for the construction of novel gravity theories. In consequence, all
solutions of these theories will obey the acceleration bounds. It is clear that
this might not be enough to get rid of all curvature singularities, but, in
any case, this is an important first step of improvement on classical general
relativity.

Throughout this article, we use the geometrical setting of a \( d \)-dimensional
torsion-free manifold \( (M,g) \) with a symmetric metric \( g \) and the
compatible metric connection \( \nabla  \) which satisfies \( \nabla g=0 \).
Consider a congruence \( X \) of affinely parameterised geodesics such that
\( \nabla _{X}X=0 \) and a connecting vector field \( Y \) with \( [X,Y]=0 \).
Then the acceleration, in the direction \( Y \), between nearby geodesics of
the congruence \( X \) is measured by the geodesic deviation equation, or Jacobi
equation, 
\begin{equation}
\nabla _{X}\nabla _{X}Y=R(X,Y)X\, ,
\end{equation}
 where \( R(X,Y)Z=\nabla _{X}\nabla _{Y}Z-\nabla _{Y}\nabla _{X}Z-\nabla _{[X,Y]}Z \)
defines the Riemann tensor for all vector fields \( X,Y,Z \). The projection
of the above acceleration onto the connecting vector \( Y \) is given by the
Riemann-Christoffel tensor, which carries all its indices downstairs, since
\begin{equation}
g(\nabla _{X}\nabla _{X}Y,Y)=g(R(X,Y)X,Y)=R(X,Y,X,Y)\, .
\end{equation}
 In order to place a bound on this physically meaningful quantity, we have to
normalise the expression. There are essentially two ways of doing so. The first
is simply to divide out the squared lengths of the vectors \( X \) and \( Y \),
i.e., to normalise by \( g(X,X)g(Y,Y) \). Another possibility is to divide
by the square of the area of the parallelogram spanned by \( X \) and \( Y \),
i.e., by \( g(X,X)g(Y,Y)-g(X,Y)^{2} \). It is the latter possibility that we
will choose because it relates the projection of the geodesic deviation acceleration
to the concept of sectional curvatures of the manifold.

We will consider the kinematics of bounded sectional curvatures, in Riemannian
and Lorentzian geometries in turn, in sections \ref{sec. Riemannian} and \ref{sec. Lorentzian}.
In section \ref{sec. Dynamics}, we will devise gravitational dynamics that
automatically enforce these bounds for all solutions of the gravitational field
equations. The crucial requirement that the theory should be free of ghosts
narrows down the admissible gravitational actions to those with a Dirac-Born-Infeld
invariance group, as will be explained in section \ref{sec, DBI}. Highly interesting
applications of the new framework, especially to the case of static, spherically
symmetric spacetimes, will be derived in section \ref{sec. Applications}. There,
we will find that the kinematics of bounded sectional curvatures imply drastic
changes in the spacetime structure in comparison to the Schwarzschild solution.
In particular, spacelike singularities will be excluded. We conclude with a
discussion in section \ref{sec. Discussion}.

\section{Kinematics of Bounded Sectional Curvatures}

\subsection{Riemannian Geometry\label{sec. Riemannian}}

Consider a \( d \)-dimensional Riemannian spacetime \( (M,g) \) with a symmetric
positive definite metric \( g \). The sectional curvature \( S_{p}(X,Y) \)
at a point \( p\in M \) with respect to a plane spanned by two vectors \( X,Y\in T_{p}M \)
is the total curvature of the two-surface of geodesics through \( p \) and
tangent to that plane. It is induced from the Riemann-Christoffel tensor by
\begin{equation}
\label{eq. defsec}
S_{p}(X,Y)\equiv \frac{R(X,Y,X,Y)}{G(X,Y,X,Y)}\, ,
\end{equation}
 where the tensor \( G:\, (T_{p}M)^{4}\rightarrow \mathbb {R} \) is defined
by 
\begin{equation}
\label{eq. Ddef}
G(X,Y,Z,W)\equiv g(X,Z)g(Y,W)-g(X,W)g(Z,Y)\, .
\end{equation}
 Note that \( G(X,Y,X,Y) \) is the squared area of the parallelogram formed
by \( X \) and \( Y \), so that the sectional curvature \( S_{p}(X,Y) \)
determines exactly the normalised value of the projection of the geodesic deviation
acceleration on the connecting vector, as it has been discussed in the introduction.
The tensor \( G \) has the same symmetry properties as the Riemann-Christoffel
tensor and the squared parallelogram area \( G(X,Y,X,Y) \) is always positive
as long as \( X \) and \( Y \) are linearly independent. For later use, we
define the set 
\begin{equation}
W_{p}\equiv \left\{ (X,Y)\in (T_{p}M)^{2}\, |\, X,Y\, \textrm{linearly independent}\right\} .
\end{equation}

The sectional curvature is a function on the Grassmannian of two-planes at the
point \( p \), i.e., \( S_{p}:\, G_{2}(T_{p}M)\rightarrow \mathbb {R} \).
This follows from the fact that, under a non-singular change of basis from \( X,Y \)
to \( X',Y' \), according to\begin{subequations}
\begin{eqnarray}
X' & = & aX+bY\, ,\\
Y' & = & cX+dY\, ,
\end{eqnarray}
\end{subequations}which is a \( GL(2,\mathbb {R}) \)-transformation with non-vanishing
determinant \( (ad-bc)\neq 0 \), both \( R(X,Y,X,Y) \) and \( G(X,Y,X,Y) \)
are multiplied by the same positive factor \( (ad-bc)^{2} \). Thus, the value
of the ratio \( S_{p}(X',Y')=S_{p}(X,Y) \) stays unchanged. The Grassmannian
\( G_{2}(T_{p}M) \) can be understood as the set of pairs of linearly independent
vectors \( W_{p} \) with its elements identified under the equivalence relation
of a \( GL(2,\mathbb {R}) \) basis change, \( G_{2}(T_{p}M)\cong W_{p}/GL(2,\mathbb {R}) \).
Note that the Grassmannian is not a vector space, but a variety. We will explain
in section \ref{sec. Lorentzian} why this is a point of crucial importance
when studying Lorentzian manifolds. A further important fact is that all sectional
curvatures at a point \( p \) determine the Riemann tensor there uniquely \cite{CCL98}.

Now we establish a bound on the absolute value of all sectional curvatures and
require 
\begin{equation}
\label{eq. bound}
\left| S(X,Y)\right| \leq \lambda ^{-1}
\end{equation}
 for all linearly independent \( (X,Y)\in W_{p} \) and for all \( p\in M \)
(dropping the subscript \( p \)). The constant \( \lambda  \) has the dimension
length squared. Therefore, any theory of gravity realising such a bound would
involve a fundamental, very short length scale at which considerable differences
to known gravity solutions could be expected. The requirement of bounded sectional
curvature can be rewritten in the form 
\begin{equation}
\label{eq. Hbound}
H^{\pm [ab]}_{\quad \, \, \, \, \, [cd]}X_{[a}Y_{b]}X^{[c}Y^{d]}\geq 0\, .
\end{equation}
 This condition has to be satisfied for both signs, which arise from the removal
of the absolute value, and the \( H^{\pm } \) are defined by 
\begin{equation}
\label{eq. endor}
H^{\pm [ab]}_{\quad \, \, \, \, \, [cd]}=\delta ^{[ab]}_{\quad [cd]}\pm R^{[ab]}_{\quad [cd]}\, ,
\end{equation}
 where \( \delta  \) arises from \( G \) by raising two indices. The \( H^{\pm } \)
are linear maps \( H^{\pm }:\, (T_{p}M)^{\otimes 2}\rightarrow \bigwedge ^{2}T_{p}M \).
Due to the symmetries of \( G \) and \( R \), they map any two-tensor into
an antisymmetric two-tensor. For the discussion of the condition on the sectional
curvatures, however, we can restrict \( H^{\pm } \) to endomorphisms of antisymmetric
two-tensors and consider \( H^{\pm }\in \textrm{End}(\bigwedge ^{2}T_{p}M) \).
The endomorphism \( \delta  \) simply becomes the identity on \( \bigwedge ^{2}T_{p}M \).

We need another definition. An element \( \Omega \in \bigwedge ^{2}T_{p}M \)
is called \textit{simple}, if it can be decomposed as \( \Omega =X\wedge Y \)
for \( X,Y\in T_{p}M \). We write 
\begin{equation}
\mathcal{Q}_{p}\equiv \left\{ \Omega \in {\bigwedge }^{2}T_{p}M\, |\, \Omega \, \textrm{simple}\right\} .
\end{equation}
 Any element of \( \bigwedge ^{2}T_{p}M \) can be written as a sum of simple
elements because the basis elements of this space can be chosen to be simple.
This allows us to consider the tensor \( G \), defined in (\ref{eq. Ddef}),
as a map \( G:\, (\bigwedge ^{2}T_{p}M)^{2}\rightarrow \mathbb {R} \) by using
the identification \( G(X\wedge Y,Z\wedge W)=G(X,Y,Z,W) \).

The endomorphisms \( H^{\pm } \) are symmetric with respect to \( G \), i.e.,
\( G(\Omega ,H^{\pm }\Sigma )=G(H^{\pm }\Omega ,\Sigma ) \) for all \( \Omega ,\Sigma \in \bigwedge ^{2}T_{p}M \).
Therefore, the \( H^{\pm } \) are diagonalisable with real eigenvalues by an
orthogonal transformation \( \mathcal{S}\in O(\textrm{dim}(\bigwedge ^{2}T_{p}M)) \)
which satisfies \( G(\mathcal{S}\Omega ,\mathcal{S}\Sigma )=G(\Omega ,\Sigma ) \).
This implies that there exists an orthonormal basis \( \{\Omega _{I}\, |\, G(\Omega _{I},\Omega _{J})=\delta _{IJ}\} \)
in \( \bigwedge ^{2}T_{p}M \) of common eigenvectors of \( H^{+} \) and \( H^{-} \)
(or, more precisely, of \( R \)) with \( H^{\pm }\Omega _{I}=\mu _{I}^{\pm }\Omega _{I} \).
In terms of this basis, we may write any plane as a linear combination, \( X\wedge Y=\sum _{I}a_{I}\Omega _{I} \).
From (\ref{eq. Hbound}), the bound on the sectional curvatures then is seen
to be equivalent to 
\begin{equation}
G(X\wedge Y,H^{\pm }(X\wedge Y))=\sum _{I}a_{I}^{2}\mu ^{\pm }_{I}\geq 0
\end{equation}
 for all \( (X,Y)\in W_{p} \) and all points of the spacetime manifold. This
is satisfied if, and only if, the eigenvalues \( \mu _{I}^{\pm } \) are non-negative,
\( \mu _{I}^{\pm }\geq 0 \). Note that this is a fully covariant requirement
because the characteristic polynomials of \( H^{\pm } \), and hence their eigenvalue
equations 
\begin{equation}
\label{eq. charpol}
\det \left( \mu \delta -H^{\pm }\right) =0\, ,
\end{equation}
 are diffeomorphism-invariant. Hence, the following holds:

\textit{An upper bound (\ref{eq. bound}) on the absolute value of all sectional
curvatures of a Riemannian manifold is equivalent to the requirements that }

\begin{lyxlist}{00.00.0000}
\item [\textit{(i)}]\noindent \textit{the endomorphisms \( H^{\pm }\in \textrm{End}(\bigwedge ^{2}T_{p}M) \)
defined in (\ref{eq. endor}) should be diagonalisable by an element \( \mathcal{S} \)
of the local orthogonal group \( O(\textrm{dim}(\bigwedge ^{2}T_{p}M)) \),
where \( \textrm{dim}(\bigwedge ^{2}T_{p}M)=d(d-1)/2 \), such that \( \mathcal{S}^{-1}H^{\pm }\mathcal{S}=\textrm{diag}(\mu _{I}^{\pm }) \)
, and }
\item [\textit{(ii)}]\noindent \textit{\( H^{\pm } \) should have non-negative eigenvalues}
\( \mu _{I}^{\pm }\geq 0 \)\textit{.}
\end{lyxlist}
The first requirement is, of course, redundant in the Riemannian setting. But
the formulation used above lends itself much better to a generalisation to the
Lorentzian case than the simpler statement that \( H^{\pm } \) should be positive
semi-definite.

\subsection{Lorentzian Geometry\label{sec. Lorentzian}}

In the Lorentzian case, the spacetime metric is indefinite with signature \( (1,d-1) \).
Consequently, the squared area \( G(X,Y,X,Y) \) in the definition of sectional
curvature in (\ref{eq. defsec}) may have either sign or even vanish. Accordingly,
a plane spanned by the basis vectors \( X,Y\in T_{p}M \) is called timelike
when \( G(X,Y,X,Y)<0 \), spacelike for \( G(X,Y,X,Y)>0 \) and null for \( G(X,Y,X,Y)=0 \).
Although the sectional curvature thus is only defined on non-null planes, the
knowledge of it on all these planes still determines the Riemann tensor uniquely.

Bounding the sectional curvatures on all non-null planes by condition (\ref{eq. bound})
is equivalent to requiring that\begin{subequations}\label{eq. condz}
\begin{eqnarray}
G(X\wedge Y,H^{\pm }(X\wedge Y))\geq 0 & \, \textrm{for}\,  & G(X\wedge Y,X\wedge Y)>0\, ,\\
G(X\wedge Y,H^{\pm }(X\wedge Y))\leq 0 & \, \textrm{for}\,  & G(X\wedge Y,X\wedge Y)<0
\end{eqnarray}
\end{subequations}for all \( (X,Y)\in W_{p} \) and for all \( p\in M \). Again
these conditions have to be satisfied for both signs. Now comes an important
point! While in the Riemannian case a large class of manifolds possesses bounded
sectional curvatures, it turns out that imposing such bounds in the Lorentzian
case is extremely restrictive: on null planes, the Riemann curvature has to
vanish, \( R(X,Y,X,Y)=0 \). Otherwise the sectional curvature could not be
bounded for nearby planes \cite{ONe1983}. But then it follows that
the sectional curvature at any given point of the manifold has the same value
for every non-null plane \cite{DaNo80}; hence, the whole manifold is one of
constant curvature \cite{GKV02}. Clearly, this is not desirable for a theory
of gravity as it could never cover a large enough range of solutions! These
arguments show that a theory of gravity on a Lorentzian spacetime manifold cannot
have bounds on \textit{all} sectional curvatures. But the same conclusion, that
the manifold has constant curvature, already follows from certain weaker assumptions,
namely from bounds on either all timelike or all spacelike planes, or from either
an upper or a lower bound on all non-null planes \cite{GKV02,GrNo78,Kul79}.

Thus, in order to avoid these rigidity theorems, a less restrictive condition
has to be devised. Such a condition amounts to a selection of planes for which
the sectional curvatures should be bounded. Observe that the above restriction
of the sectional curvature map \( S:\, G_{2}(T_{p}M)\rightarrow \mathbb {R} \)
to all non-null (timelike, spacelike) planes in the Grassmannian \textit{variety}
\( G_{2}(T_{p}M) \) is not a natural one from the point of view of algebraic
geometry, as these sets are subsets but not \textit{subvarieties} of \( G_{2}(T_{p}M) \).
A homomorphism between varieties, however, can only be sensibly restricted to
subvarieties. In the following, we will therefore identify the maximal subvariety
contained in the set of timelike (spacelike) planes and only constrain the sectional
curvature for these planes. More precisely, consider the vector space \( \bigwedge ^{2}T_{p}M \)
which has already played an important r\^{ o}le in the Riemannian construction.
Not all elements \( \Omega  \) of that space describe planes, but only the
simple ones, \( \Omega \in \mathcal{Q}_{p} \). Each plane corresponds to a
pair of linearly independent basis vectors, i.e., to an element of \( W_{p} \).
But \( W_{p}/SL(2,\mathbb {R}) \) is isomorphic to \( \mathcal{Q}_{p}\setminus \{0\} \)
via \( (X,Y)\mapsto X\wedge Y \). So \( \mathcal{Q}_{p} \) additionally contains
a degenerate plane for which \( X \) becomes parallel to \( Y \). The set
of planes \( \mathcal{Q}_{p} \) forms a variety embedded in the vector space
\( \bigwedge ^{2}T_{p}M \), for its elements satisfy the polynomial equation
\( \Omega \wedge \Omega =0 \), compare e.g. \cite{Har92}. (If the length of
the basis vectors is projected out, one gets the Pl\"{ u}cker embedding \( G_{2}(T_{p}M) \hookrightarrow \mathbb {P}(\bigwedge ^{2}T_{p}M) \).)
A mathematically well-motivated subset of planes, on which one might hope to
achieve bounded sectional curvatures in Lorentzian manifolds, therefore, should
be a subvariety of the set of all planes \( \mathcal{Q}_{p} \) and not, for
instance, the subset of all timelike planes which is not a subvariety since
\( G(X,Y,X,Y)<0 \) is not a polynomial equation.

We return to the conditions (\ref{eq. condz}) for bounded sectional curvature
on all non-null planes. The set of timelike planes for which the conditions
hold can be written as the intersection 
\begin{equation}
\label{eq. intsec}
\mathcal{Q}_{p}\cap \left\{ \Omega \in {\bigwedge }^{2}T_{p}M\, |\, G(\Omega ,\Omega )<0,\, G(\Omega ,H^{\pm }\Omega )\leq 0\right\} ,
\end{equation}
 and in an analogous form with reversed relations for the spacelike planes.
This set is not a subvariety of \( \mathcal{Q}_{p} \), as it is not defined
by additional polynomial equations. But into it, we can embed a `maximal subvariety'
by constructing the vector space of maximal dimension which can be embedded
into the set in curly brackets. The intersection of such a linear subspace with
\( \mathcal{Q}_{p} \) is then the desired subvariety. We show now that this
vector space can be found, in the present Lorentzian case, by requiring the
conditions \textit{(i)} and \textit{(ii)} on the endomorphisms \( H^{\pm } \)
which we have established in the Riemannian case.

By \textit{(i)}, the endomorphisms \( H^{\pm } \) should be diagonalisable
by the local orthogonal group which is now \( O(d-1,(d-1)(d-2)/2) \). This
requirement is non-trivial in the Lorentzian case: the \( H^{\pm } \) are still
symmetric, but now with respect to an indefinite metric, which does not guarantee
diagonalisability. Requiring diagonalisability, now, implies the existence of
an orthonormal basis 
\begin{equation}
\{\Omega _{I},\Omega _{\widetilde{I}}\, |\, I=1\, ...\, (d-1),\, \widetilde{I}=1\, ...\, (d-1)(d-2)/2\}
\end{equation}
 of \( \bigwedge ^{2}T_{p}M \) with \( G(\Omega _{I},\Omega _{J})=-\delta _{IJ} \),
\( G(\Omega _{\widetilde{I}},\Omega _{\widetilde{J}})=\delta _{\widetilde{I}\widetilde{J}} \)
and \( G(\Omega _{I},\Omega _{\widetilde{J}})=0 \) which consists of common
eigenvectors of \( H^{\pm } \) with corresponding eigenvalues \( \{\mu ^{\pm }_{I},\mu ^{\pm }_{\widetilde{I}}\} \).
And by \textit{(ii)}, these eigenvalues \( \mu ^{\pm }_{I} \) and \( \mu ^{\pm }_{\widetilde{I}} \)
should be non-negative.

Both these requirements together imply that the linear span of timelike eigenvectors
\begin{equation}
\label{eq. span}
\left\langle \Omega _{I}\, |\, I=1\, ...\, (d-1)\right\rangle 
\end{equation}
 is the sought-for highest-dimensional linear subspace of \( \bigwedge ^{2}T_{p}M \)
which can be embedded into the set in curly brackets appearing in equation (\ref{eq. intsec}).
This is most easily seen from the decomposition \( \Omega =\sum _{I}a_{I}\Omega _{I}+\sum _{\widetilde{I}}a_{\widetilde{I}}\Omega _{\widetilde{I}} \),
which gives 
\begin{equation}
G(\Omega ,H^{\pm }\Omega )=-\sum _{I}a_{I}^{2}\mu ^{\pm }_{I}+\sum _{\widetilde{I}}a_{\widetilde{I}}^{2}\mu ^{\pm }_{\widetilde{I}}\, .
\end{equation}
 For elements \( \Omega  \) of the linear span (\ref{eq. span}) of the time-like
eigenvectors, the above expression is clearly negative. Another basis vector
that could be added to enlarge the dimension of this vector space would necessarily
be spacelike, so that the augmented space could not be properly embedded anymore.
An analogous argument where the spacelike eigenvectors \( \Omega _{\widetilde{I}} \)
replace the timelike eigenvectors \( \Omega _{I} \) holds for the case of the
spacelike planes. Thus we have shown the following.

\textit{On a Lorentzian manifold, the requirements (i) and (ii) on the endomorphisms
\( H^{\pm } \), as formulated at the end of the preceding subsection, imply
bounds on the sectional curvatures of certain subvarieties of the variety \( \mathcal{Q} \)
of all planes in \( \bigwedge ^{2}TM \). In terms of the common eigenvector
basis \( \{\Omega _{I},\Omega _{\widetilde{I}}\} \) of \( H^{\pm } \), these
subvarieties are given by the timelike planes in \( \left\langle \Omega _{I}\right\rangle \cap \mathcal{Q} \)
and by the spacelike planes in \( \left\langle \Omega _{\widetilde{I}}\right\rangle \cap \mathcal{Q} \). }

Note that this mechanism of choosing the planes of bounded sectional curvature
is a dynamical one, as it is the Riemann tensor (of particular spacetime solutions)
that determines the eigenvectors of the endomorphisms \( H^{\pm } \). This
construction, bounding the sectional curvatures on certain subvarieties of the
variety of all planes hence allows to circumvent the rigidity theorems obtained
by Nomizu and others.

\section{Gravitational Dynamics\label{sec. Dynamics}}

We have now established algebraic conditions that allow placing sensible bounds
on the sectional curvatures along certain planes. But how does one implement
that solutions of gravitational equations, yet to be devised, automatically
give rise to non-negative eigenvalues for the endomorphisms \( H^{\pm } \)?
A fairly general answer to this question is to use, in the equations of motion,
absolutely converging power series in the Riemann tensor that converge on a
domain with non-negative eigenvalues. For definiteness, consider the square
roots 
\begin{equation}
\sqrt{H^{\pm }}^{A}_{\, \, B}=\sqrt{\delta \pm \lambda R}^{A}_{\, \, B}\, ,
\end{equation}
 where we use the Petrov notation for indices of \( \bigwedge ^{2}T_{p}M \).
The index \( A\sim [ab] \) may take \( d(d-1)/2 \) values and corresponds
to a pair of antisymmetrised indices of \( T_{p}M \). If the above expressions
appear in the equations of motion, then any diagonalisable solution will
force the eigenvalues of \( H^{\pm } \) to be real and non-negative. Otherwise the solution
will become complex.

It is very convenient, for a number of reasons, to derive the gravitational
dynamics from an action principle. Not only does this allow to check easily
whether the modified theory of gravity is consistent with Einstein's theory
in a certain limit, which is sufficient for its compatibility with experiment,
it also simplifies the discussion and exclusion of ghosts. Gravitational ghosts
are modes of negative kinetic energy in a perturbative expansion of the metric
around the Minkowski vacuum. Ghosts appear whenever there are quadratic terms
in the curvature expansion of the action; the only exception from this rule
is the quadratic Gauss-Bonnet combination 
\begin{equation}
\sqrt{-g}\left( R^{2}-4R_{ab}^{2}+R_{abcd}^{2}\right) \, ,
\end{equation}
 which is a total derivative in four dimensions, and for which, also in higher
dimensions, the kinetic terms of the perturbative expansion cancel.

As a consequence of the kinematical construction above, the Lagrangian for an
action implying partially bounded sectional curvatures should be constructed
from a power series \( \sum _{n=0}^{\infty }a_{n}(\lambda R)^{n\, A}_{\quad B} \)
with the desired finite convergence radius to enforce non-negative eigenvalues
for \( H^{\pm } \). Furthermore, the theory should be ghost-free. It is natural
to use the same power series for \( H^{+} \) and \( H^{-} \), i.e., for \( \pm \lambda  \),
in constructing the dynamics. Hence, the Lagrangian should be built from the
expression 
\begin{equation}
\sum _{n=0}^{\infty }a_{n}(\lambda R)^{n\, A}_{\quad B}-\sum _{n=0}^{\infty }a_{n}(-\lambda R)^{n\, A}_{\quad B}\, ,
\end{equation}
 where the minus sign between the sums is dictated by the requirement of ghost-freedom.
This is the only possible way to achieve ghost-freedom in the proposed setup:
contractions of the Riemann tensor in Petrov notation can never yield the Ricci
tensor which appears in the Gauss-Bonnet term. Note that the
cancellation of the ghosts
automatically kills an otherwise possible cosmological constant term which,
however, could again be included by hand without inconsistency. Because of their
absolute convergence, the two series may be rearranged, and they collapse to
\begin{equation}
\sum _{n=0}^{\infty }a_{2n+1}(\lambda R)^{(2n+1)\, A}_{\quad \qquad B}=a_{1}\lambda R^{A}_{\, \, B}+\mathcal{O}(\lambda ^{3})\, .
\end{equation}
 The simplest way to build a scalar action from the above expression is to act
on it with the linear integral-trace functional, to obtain 
\begin{equation}
\label{eq. genaction}
\int \sqrt{-g}\, \, \, \textrm{Tr}\sum _{n\, \textrm{odd}}a_{n}(\lambda R)^{n\, A}_{\quad B}\, .
\end{equation}
 This is then seen to be a direct generalisation of Einstein's
theory, where \( a_{2n+1}=0 \) for \( n>0 \). So one particular action whose
solutions by construction have bounded sectional curvature (on all planes in
the Riemannian case, and on certain subvarieties in the Lorentzian case) could
be taken as 
\begin{equation}
S_{\textrm{example}}=\frac{1}{\lambda }\, \int \sqrt{-g}\, \, \, \textrm{Tr}\left( \sqrt{H^{+}}^{A}_{\, \, B}-\sqrt{H^{-}}^{A}_{\, \, B}\right) .
\end{equation}
 The equations of motion are derived in the usual manner by variation with respect
to the spacetime metric \( g_{ab} \): they present a system of fourth-order
partial differential equations. In the limit \( \lambda \rightarrow 0 \), this
action reproduces the Einstein-Hilbert action up to a constant that can be divided
out. The same limit simplifies the equations of motion to yield Einstein's vacuum
equations \( R_{ab}-\frac{1}{2}Rg_{ab}=0 \).

Although the action (\ref{eq. genaction}) appears to be the mathematically
most straightforward one, we could have made several alternative choices in
the construction. Another obvious choice would have been to take a determinant
instead of a trace. It is also possible first to make scalars out of the power
series for \( \pm \lambda  \) and then to subtract. This would lead, for instance,
to the action 
\begin{equation}
\int \sqrt{-g}\left( \det \sqrt{H^{+}}-\det \sqrt{H^{-}}\right) \, .
\end{equation}
 The first part of this expression is almost of the form as has been considered
in \cite{Woh04}. There are, however, two important differences here. Firstly,
the inclusion of the second term bounds the value of the sectional curvatures
(on the planes that can be linearly combined solely from spacelike or timelike
eigenvectors of \( H^{\pm } \)) additionally from below, and, notably, this
removes the ghosts as well. Secondly, it is important to take the square root
of the matrix first and then the determinant, because this is what constrains
the eigenvalues of \( H^{\pm } \) when only real solutions are considered.

\section{Dirac-Born-Infeld Invariance\label{sec, DBI}}

Our construction of gravity theories with partially bounded sectional curvatures
features a fundamental length scale, appearing in the constant \( \lambda  \)
which has the dimension length squared. Hence, one expects that gravitational
dynamics respecting that scale should lead to solutions very different from
those of general relativity, especially on scales which are small compared to
\( \sqrt{\lambda } \) . Such a crucial change in the short distance behaviour
should also be mirrored in structural differences. To show that this is indeed
the case, we need to briefly review the kinematics associated to Dirac-Born-Infeld theory.

The singularities of Maxwell electrodynamics, most prominently the divergence
of the electric field energy density of a point charge, prompted Born and Infeld
in \cite{BoIn34} to devise a nonlinear theory of electrodynamics, which regulates
this divergence by the introduction of a parameter \( \lambda  \) of dimension
length squared. In string theory, the corresponding action emerges \cite{FrTs85}
as the low energy effective action of a \( U(1) \) gauge field \( A \), with
corresponding field strength \( F=dA \), on a \( Dp \)-brane \( \Sigma  \),
\begin{equation}
\label{DBI}
\int _{\Sigma }\sqrt{-g}\sqrt{\det (\delta +\lambda F)}\, .
\end{equation}
 This theory is manifestly invariant under worldvolume diffeomorphisms of the
\( (p+1) \)-dimensional Lorentzian manifold \( \Sigma  \), and hence possesses
a local \( O(1,p) \) Lorentz symmetry. The feature of a finite length scale
in (\ref{DBI}), however, has profound structural consequences. Based on the
observation that this action only depends on even powers of the electromagnetic
field strength \( F \), it has been shown in \cite{ScPf03} that there is a
hidden dynamical invariance, extending the local invariance group to a product
group, 
\begin{equation}
\label{embed}
O(1,p)\hookrightarrow O(1,p)\times O(1,p)\, ,
\end{equation}
 into which the Lorentz symmetry group is diagonally embedded. An investigation
of the kinematical meaning of the product group, acting on curves in the frame
bundle, shows that it contains, apart from the standard Lorentz transformations,
transformations to arbitrarily rotating and accelerated frames, whose covariant
acceleration for non-rotating observers is bounded by the inverse of the length
scale, \( 1/\sqrt{\lambda } \). As much as the extension of the rotation group
\( O(3) \) to \( O(1,3) \) captures the dynamical symmetries of Maxwell theory
due to the existence of a fundamental speed, the extension (\ref{embed}) is
implied by the existence of a fundamental squared length scale \( \lambda  \)
in Dirac-Born-Infeld theory. That the extended group also captures the regulation
of Born-Infeld theory is further illustrated by the following fact: if one defines
particles in quantum field theory as the irreducible representations of a corresponding
doubling of the Poincare group, then every particle automatically is accompanied
by a Pauli-Villars regulating Weyl ghost \cite{SWG03}.

These insights are intimately linked to the findings of the present paper: in
the previous chapter, we have seen that the criterion of ghost-freedom for gravity
theories with partially bounded sectional curvatures leads to actions of the
type (\ref{eq. genaction}), to which solely odd powers of the Riemann tensor,
viewed as an endomorphism on \( \bigwedge ^{2}T_{p}M \), contribute. Hence
one can show in close technical analogy to \cite{ScPf03} that gravitational
ghost-freedom can be understood as singling out those theories that possess
a hidden invariance (\ref{embed}). In turn, the extended structure group \( O(1,p)\times O(1,p) \)
appears not only to capture the regulation of Born-Infeld theory, but also the
absence of ghosts there, and transfers this property to curvature regulated
gravity theories in our sense.

\textit{In summary, we arrive at the remarkable fact that ghost-free gravity
theories with tidal acceleration bounds, as constructed in the preceding sections,
automatically imply kinematics that put an upper bound of the same value \( 1/\sqrt{\lambda } \)
on covariant accelerations as well. }

These structural insights justify the classification of the gravity theories
presented in this paper as of `Born-Infeld type'.

\section{Simple Applications\label{sec. Applications}}

The appeal of the geometrical construction scheme presented here is the characterisation
of a spacetime with partially bounded sectional curvatures, and hence, partially
bounded geodesic deviation accelerations, by a simple algebraic property of
the Riemann tensor: the endomorphisms \( H^{\pm } \) should be diagonalisable
and have real non-negative eigenvalues. Even without considering specific gravitational
dynamics this has interesting consequences, valid for the whole class of theories
constructed here. Two applications will be considered: for asymptotically flat,
static, spherically symmetric spacetimes we will show the non-existence of spacelike
singularities, and a discussion of pp-wave backgrounds will illustrate some
limitations of the proposed scheme.

\subsection{pp-Wave Spacetimes}

The first application considers pp-wave spacetimes (plane-fronted waves with
parallel rays) which are defined by the existence of a covariantly constant
null Killing vector field \( \partial /\partial v \). The line element can
be written in the form 
\begin{equation}
ds^{2}=-f(u,x^{i})du^{2}-2dudv+d\mathbf{x}\cdot d\mathbf{x}\, ,
\end{equation}
 and it can be shown that in these spacetimes all curvature invariants vanish
\cite{Pla}. This holds, in particular, for the invariants 
\begin{equation}
\textrm{Tr}\left[ R^{n\, A}_{\quad B}\right] ,\quad \textrm{for}\, \, n\in \mathbb {N}\, ,
\end{equation}
 which determine the eigenvalues of \( H^{\pm } \) since they appear as the
coefficients of the characteristic polynomial in equation (\ref{eq. charpol}).
For the pp-wave spacetimes, the eigenvalue equation becomes \( (\mu -1)^{d(d-1)/2}=0 \),
and so all eigenvalues equal \( \mu =1 \), as is the case for Minkowski space.
This also means that the Riemann tensor, although it is non-trivial, is nilpotent.
More explicitly, in four dimensions and in Petrov notation using the labelling
\( ([uv],[ux^{1}],[ux^{2}],[vx^{1}],[vx^{2}],[x^{1}x^{2}])\equiv (1,\, ...\, ,6) \),
one finds the nonzero components 
\begin{equation}
\left( \begin{array}{cc}
R^{4}_{\, \, 2} & R^{4}_{\, \, 3}\\
R^{5}_{\, \, 2} & R^{5}_{\, \, 3}
\end{array}\right) =\frac{1}{2}\left( \begin{array}{cc}
\partial _{x^{1}}^{2}F & \partial _{x^{1}}\partial _{x^{2}}F\\
\partial _{x^{1}}\partial _{x^{2}}F & \partial _{x^{2}}^{2}F
\end{array}\right) .
\end{equation}
 This Riemann tensor is not diagonalisable by a local \( O(3,3) \) transformation,
since it does not admit an eigenvector basis for \( \bigwedge ^{2}T_{p}M \).

But because the Riemann tensor is nilpotent with \( R^{2\, A}_{\quad B}=0 \),
the pp-waves are solutions of any theory constructed within our framework, as
they solve Einsteins theory on the remaining first order level of the curvature
expansion, compare (\ref{eq. genaction}). The polarisation of the waves may
diverge and lead to `null-singularities'. These singularities appear in the
tidal forces given by the geodesic deviation equation and have been argued to
be unavoidable in any theory of gravity whose Lagrangian depends only on the
metric, covariant derivatives and the Riemann tensor \cite{HoMy95}. In particular,
our construction does not escape that conclusion.

\subsection{Static Spherical Symmetry}

The second, particularly nice example considers static, spherically symmetric
solutions in four dimensions with a spacetime ansatz of the form 
\begin{equation}
ds^{2}=-A(r)dt^{2}+B(r)dr^{2}+r^{2}d\Omega _{2}^{2}\, ,
\end{equation}
 where \( A \) and \( B \) are arbitrary functions of the radial coordinate
\( r \) and \( d\Omega _{2}^{2} \) gives the standard line element on the
unit two-sphere.

Now we establish sectional curvature bounds in the sense of section \ref{sec. Lorentzian},
by assuming that the endomorphisms \( H^{\pm } \), which are already diagonal
in the adjoint basis induced by the Schwarzschild coordinates, have non-negative
eigenvalues. One of these eigenvalues is algebraic in \( B \). The corresponding
condition is 
\begin{equation}
1\pm \lambda \frac{1-B(r)}{r^{2}B(r)}\geq 0
\end{equation}
 and implies that \( B \) is bounded from below for all \( r \) by the function
\( B_{\textrm{low}}(r)=\lambda /(\lambda +r^{2}) \) and from above, for \( r<\sqrt{\lambda } \),
by \( B_{\textrm{up}}(r)=\lambda /(\lambda -r^{2}) \). In particular, this
implies that \( B \) is strictly positive everywhere and that \( B(0)=1 \)
with slope \( B'(0)=0 \). An argument given in \cite{Hol02} shows that such
behaviour of the function \( B \) is necessary in any theory of gravity on
a Lorentzian spacetime manifold that regulates the Schwarzschild singularity.
The framework presented here obviously passes this test.

When the gravitational equations of motion are constructed from an action principle,
there is another implicit requirement to satisfy. The density \( \sqrt{-g}=r^{2}\sqrt{AB} \)
in the integrand is only well-defined when 
\begin{equation}
A(r)B(r)\geq 0
\end{equation}
 everywhere. But as \( B(r)>0 \) this shows \( A(r)\geq 0 \).

Already at this point, we see that theories with partially bounded sectional
curvatures show a completely different behaviour than does general relativity,
where, by Birkhoff's theorem, the Schwarzschild solution is the unique static,
spherically symmetric and asymptotically flat vacuum solution, with \( A=B^{-1}=1-2m/r \)
for a mass parameter \( m \). For positive mass \( m>0 \), the Schwarzschild
spacetime has a horizon at \( r=2m \), in whose interior \( A \) and \( B \)
become negative. All timelike and null geodesics are trapped behind this horizon,
and there is a spacelike singularity at \( r=0 \). In our theories with partially
bounded sectional curvature, however, a change of signs in the functions \( A \)
and \( B \) is forbidden. Hence, hypersurfaces of constant \( r \) can never
be spacelike, such that spacelike singularities are excluded from all possible
solutions. (Singularities can only occur at constant radius due to spherical
symmetry.) This mechanism does not prevent the occurrence of timelike singularities
or, as seen previously, null singularities.

\begin{figure}
\resizebox*{!}{0.3\textheight}{\includegraphics{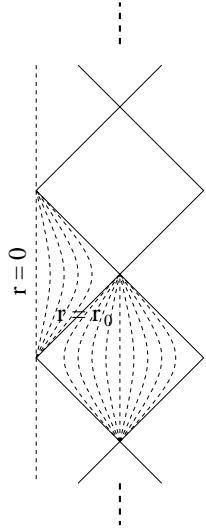} }

\caption{\textit{\label{fig.Penrose}The Carter-Penrose diagram of static, spherically
symmetric, asymptotically flat spacetimes with \protect\protect\( A,B\protect \protect \)
finite and strictly positive everywhere on \protect\protect\( r\geq 0\protect \protect \)
except \protect\protect\( A(r_{0})=0\protect \protect \) demonstrates the possibility
of horizons in theories with partially bounded sectional curvatures. Several
asymptotic regions with their own future and past infinities occur. A few surfaces
of constant radius are indicated.}}
\end{figure}

Horizons may still appear in solutions. To see this more explicitly consider
the spherically symmetric spacetimes in which the functions \( A \) and \( B \)
are finite for all \( r\geq 0 \). When \( A>0 \), i.e., when \( A \) is strictly
positive everywhere, nothing interesting happens. There are no horizons and
the spacetime is conformally equivalent to Minkowski space. Now assume that
\( A \) vanishes at an isolated point, \( A(r_{0})=0 \). In the usual manner,
it is possible to introduce null coordinates by setting 
\begin{equation}
\tilde{r}(r)=\int ^{r}dr'\left( \frac{B(r')}{A(r')}\right) ^{1/2}
\end{equation}
 and defining \( u=t-\tilde{r} \) and \( v=t+\tilde{r} \), where \( u \)
and \( v \) are constant on outgoing or ingoing radial null geodesics, respectively.
These coordinates allow the continuation of the radial null geodesics across
\( r=r_{0} \) because it takes a finite affine parameter distance \( \tau _{10} \)
to reach this hypersurface from a point \( r_{1}>r_{0} \), 
\begin{equation}
\tau _{10}\sim \int _{r_{0}}^{r_{1}}dr\left( A(r)B(r)\right) ^{1/2}\, .
\end{equation}
 This argument extends and shows the existence of a horizon at \( r=r_{0} \)
which, however, is permeable from both sides. The Carter-Penrose diagram of
the resulting spacetime is shown in figure \ref{fig.Penrose}.

Finally, let us take a closer look at the planes on which the sectional curvatures
are bounded. We use the Petrov notation and the labelling \( ([tr],[t\theta ],[t\phi ],[\theta \phi ],[\phi r],[r\theta ])=(1,\, ...\, ,6) \).
The six-vectors \( Z^{A}=(\mathbf{Z}_{1},\mathbf{Z}_{2}) \) that describe planes
then are those which satisfy \( \mathbf{Z}_{1}\cdot \mathbf{Z}_{2}=0 \). This
follows from the choice of basis \( X=(X_{0},\mathbf{X}) \) and \( Y=(Y_{0},\mathbf{Y}) \)
which implies 
\begin{equation}
\label{eq. Zform}
Z^{A}=\left( X_{0}\mathbf{Y}-Y_{0}\mathbf{X},\mathbf{X}\times \mathbf{Y}\right) \, .
\end{equation}
 The endomorphisms \( H^{\pm } \) are diagonal in the static spherically symmetric
example, as is the tensor \( G:\, (\bigwedge ^{2}T_{p}M)^{2}\rightarrow \mathbb {R} \),
for which 
\begin{equation}
G_{AB}=\textrm{diag}\left( -AB,-Ar^{2},-Ar^{2}\sin ^{2}\theta ,r^{4}\sin ^{2}\theta ,Br^{2}\sin ^{2}\theta ,Br^{2}\right) \, .
\end{equation}
 The timelike and spacelike eigenvectors of \( H^{\pm } \) are \( \Omega _{I}=(\mathbf{e}_{I},0) \)
and \( \Omega _{\tilde{I}}=(0,\mathbf{e}_{\widetilde{I}}) \), respectively.
The sectional curvature is thus bounded on the subvarieties of all timelike
planes described by \( Z^{A}=(\mathbf{Z},\mathbf{0}) \) with \( BZ_{1}^{2}+r^{2}(Z_{2}^{2}+\sin ^{2}\theta Z_{3}^{2})=1/A \)
(where the \( Z^{A} \) are normalised to avoid overcounting). More explicitly,
these planes can be described by an orthonormal basis: given any normalised
timelike vector \( X \) with \( g(X,X)=-1 \), we can calculate the second,
spacelike basis vector \( Y \) with \( g(Y,Y)=1 \) from (\ref{eq. Zform}).
When \( X=(\pm 1/\sqrt{A},\mathbf{0}) \), we find \( Y=(0,\mathbf{Y}) \) for any normalised
three-vector \( \mathbf{Y} \). For \( X_{0}\neq \pm 1/\sqrt{A} \), we find
\( Y=(Y_{0},X_{0}\mathbf{X}/Y_{0}) \) where 
\begin{equation}
Y_{0}=\pm \sqrt{\frac{AX_{0}^{2}-1}{A}}\, .
\end{equation}
 All these planes are easily visualised: they contain the local time axis \( \partial /\partial t \).
(This is obvious in the first case with \( X_{0}=\pm 1/\sqrt{A} \), while in
the second case it follows from the fact that \( \mathbf{X} \) is parallel
to \( \mathbf{Y}=X_{0}\mathbf{X}/Y_{0} \).) A similar calculation for the spacelike
planes shows that among those with bounded sectional curvature are the planes
orthogonal to the local time axis, defined by \( dt=0 \).

\section{Discussion\label{sec. Discussion}}

In contrast to past attempts to modify Einstein's theory of gravity, where the
consequences of new actions, differing from the Einstein-Hilbert action and
proposed in a more or less \textit{ad hoc} fashion, are analysed, this work
has been based on clear-cut geometrical assumptions. Motivated by bounds on
the deviation accelerations between nearby geodesics, we have investigated possible
bounds on the sectional curvatures of a spacetime, with the aim of potentially
improving on the singularity problems of classical general relativity.

In the Riemannian case, the sectional curvatures may be bounded for all planes,
whereas on the physically relevant Lorentzian manifolds, the situation is more
involved. Due to the existence of strong rigidity theorems, certain subsets
of planes have to be chosen to avoid trivial theories allowing for constant
curvature solutions only. A consistent choice of such subsets can be made by
demanding that they should form subvarieties of the variety of all possible
planes through a given point.

We have shown that the bounds on the sectional curvatures along maximal subvarieties
of the set of all planes can be reformulated in a nice algebraic way. The conditions
for these bounds translate into the requirement that certain endomorphisms,
which are constructed from the Riemann tensor and act on the space of antisymmetric
two-tensors, should be diagonalisable by the local orthogonal group and should
have non-negative eigenvalues. In addition to these kinematical results, it
has been shown that it is possible to construct ghost-free gravity actions such
that the dynamical solutions of their corresponding equations of motion automatically
satisfy the sectional curvature bounds. This is strong evidence that there exist
classical gravity theories on Lorentzian manifolds that are better-behaved than
Einstein's theory.

The strengths and limitations of the kinematical framework constructed here
have been illustrated by two instructive examples. The expectation that a regulation,
at least for some types of singularities, might be possible has been found justified.
This expectation emerged from the Dirac-Born-Infeld structure of our theories
and from earlier results in \cite{Woh04}, where actions similar to those obtained
here had been investigated. In spherically symmetric spacetimes in four dimensions,
without even considering specific dynamics, it has been shown that there cannot
be spacelike singularities, which is a very strong result, as it applies to
the whole class of curvature regulated theories in our sense. Another application
to pp-wave spacetimes explains why null singularities cannot be removed in our
theories.

A new fundamental length scale appears as an important ingredient in the construction.
We have shown that the appearance of the length scale here is intimately linked
to its r\^{o}le in Dirac-Born-Infeld theory \cite{Sch03}. The covariant acceleration
of non-geodesic observers turns out to be bounded by the same value that one
chooses to impose on tidal accelerations. It would be interesting to understand
the thereby established relations to other theories with fundamental length
scales, such as, for example, string theory or quantum gravity, which all predict
modifications of general relativity. A study of exact, or numerical, solutions
of specific gravitational dynamics constructed within our framework will allow
to constrain the possible range for the length scale parameter from comparison
with observations.

Our investigations here indicate the emergence of at least one further generic
feature besides the finite, fundamental length scale: the existence of curvature
corrections to all orders seems to be an essential ingredient in order to regulate
gravity.

\begin{acknowledgments}

The authors are grateful to Paul K. Townsend and Achim Kempf for useful discussions. FPS thanks
the Mathematical Institute at Oxford University for the opportunity to work
there. MNRW thanks the Perimeter Institute where this research was begun for
its generous hospitality. He gratefully acknowledges financial support from
the Gates Cambridge Trust.

\end{acknowledgments}

\end{document}